# THz pulses over 50 millijoules generated from relativistic picosecond laser-plasma interactions

Guoqian Liao[1,2,8], Hao Liu[1,7], Yutong Li[1,7,8,*], Graeme G. Scott[3], David Neely[3,4,*], Yihang Zhang[1,7], Baojun Zhu[1,7], Zhe Zhang[1], Chris Armstrong[4,3], Egle Zemaityte[4,3], Philip Bradford[5], Peter G. Huggard[6], Paul McKenna[4], Ceri M. Brenner[3], Nigel C. Woolsey[5], Weimin Wang[1,8], Zhengming Sheng[2,4,8,9] and Jie Zhang[2,8,*]

[1] Beijing National Laboratory for Condensed Matter Physics, Institute of Physics, Chinese Academy of Sciences, Beijing 100190, China
[2] Key Laboratory for Laser Plasmas (Ministry of Education) and School of Physics and Astronomy, Shanghai Jiao Tong University, Shanghai 200240, China
[3] Central Laser Facility, STFC Rutherford Appleton Laboratory, Didcot, OX11 0QX, UK
[4] Department of Physics SUPA, University of Strathclyde, Glasgow, G4 0NG, UK
[5] Department of Physics, York Plasma Institute, University of York, Heslington York YO10 5DD, UK
[6] Space Science and Technology Department, STFC Rutherford Appleton Laboratory, Didcot, OX11 0QX, UK
[7] School of Physical Sciences, University of Chinese Academy of Sciences, Beijing 100049, China
[8] Collaborative Innovation Center of IFSA (CICIFSA), Shanghai Jiao Tong University, Shanghai 200240, China
[9] Tsung-Dao Lee Institute, Shanghai Jiao Tong University, Shanghai 200240, China

E-mail: ytli@iphy.ac.cn, david.neely@stfc.ac.uk and jzhang1@sjtu.edu.cn

Ultrahigh-power terahertz (THz) radiation sources are essential for many applications, such as nonlinear THz physics, THz-wave based compact accelerators, *etc*. However, until now none of THz sources reported, whether based upon large-scale accelerators or high power lasers, have produced THz pulses with energies above the millijoule (mJ) barrier. Here we report on the efficient generation of low-frequency (<3 THz) THz pulses with unprecedentedly high energies over 50 mJ. The THz radiation is produced by coherent transition radiation of a picosecond laser-accelerated ultra-bright bunch of relativistic electrons from a solid target. Such high energy THz pulses can not only trigger various nonlinear dynamics in matter, but also open up a new research field of relativistic THz optics.

Terahertz (THz) radiation is usually utilized as a non-ionizing sensitive probe in many science disciplines [1,2]. With the recent advent of multi-$\mu$J-level THz pulses, THz radiation has started to serve as a mode-selective pumping driver for preparing and controlling particular transient states of matter in nonlinear THz-field-matter interactions [3,4]. THz pulses, with higher energies over the multi-mJ barrier and field strengths up to GV/m, are expected to enable more intriguing applications, including the acceleration and manipulation of charged particles [5,6], ultrafast switching or controlling of magnetic domains [7], THz-triggered surface catalytic reactions [8], and THz nano-optoelectronics *et al*. Currently, high-power THz pulses with energies up to hundreds of $\mu$J have been demonstrated from conventional electron accelerators [9−11], ultrafast laser-pumped crystals [12−14] and ultraintense femtosecond laser-produced plasmas [15,16]. The generation of THz pulses with energies over mJ, however, has remained a formidable challenge.

Here we report efficient generation of low-frequency THz pulses with unprecedentedly high energies, beyond 50 mJ, surpassing by far any other THz sources [9–22]. The THz radiation originates mainly from the coherent transition radiation induced by the huge-charge (>100 nC), high-current (>0.1 MA), energetic electron bunch produced in relativistic, picosecond (ps) laser-foil interactions.

The experiment was carried out using the Vulcan laser [23] at the Rutherford Appleton Laboratory in the UK. Figure 1 shows the experimental set-up. A high-intensity laser pulse (~1.5 ps in pulse duration) was focused by an f/3 off-axis parabola (OAP) onto a 100-$\mu$m thick Cu foil target with an incidence angle of 30° and a ~5 $\mu$m focal spot size (FWHM). For the maximum laser energy of ~60 J on target, the peak laser intensity was ~5×10$^{19}$ W/cm$^2$ ($a_0$~6). The THz radiation at 75° was collected by a pair of polymethylpentene (TPX) lens and then detected by a THz spectrometer or a THz camera. In the THz spectrometer, the THz beam was divided into 8 beamlines. Low-pass or narrowband band-pass THz filters with varying cut-off or central frequencies were inserted in the different beamlines, and the filtered THz radiation was measured by cross-calibrated pyro-electric detectors. Besides the 75° configuration shown in Fig. 1, THz radiation at 45°, −20° and −40° with respect to the rear target normal was also detected with similar configurations (not shown in Fig. 1), where "−" represents the side of laser propagation direction at the target rear. To measure the THz source size on the target rear surface, a THz imaging camera (DataRay Inc.) was used.

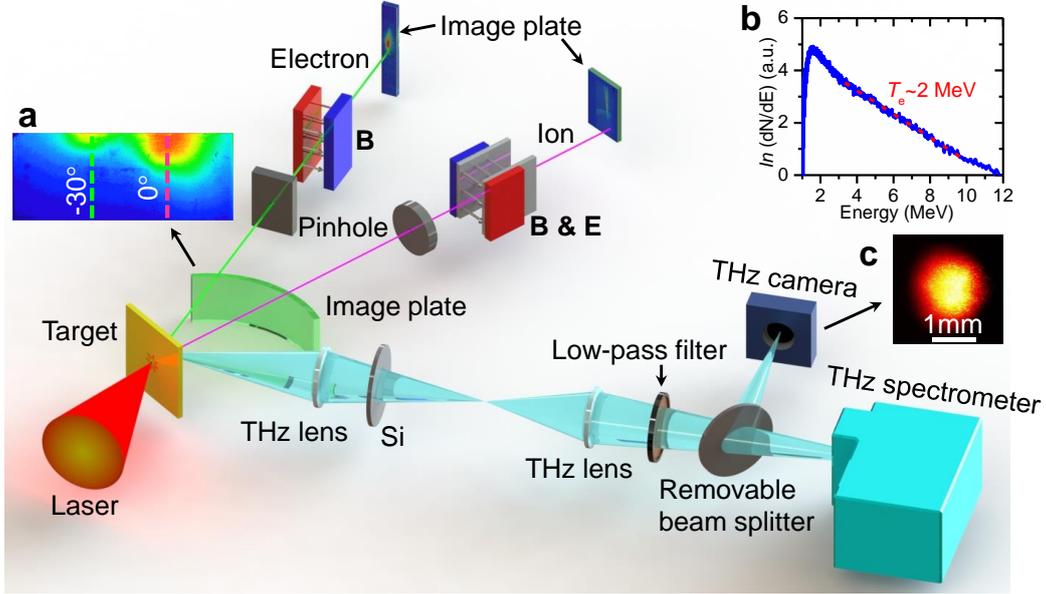

**Fig. 1.** Experimental setup**.** The THz radiation, accelerated ions and escaping electrons from the target rear were characterized simultaneously. To filter out the unwanted radiation and avoid the saturation of THz detectors, THz filters and high-resistivity silicon (Si) wafers were placed in the THz path. Inset (a), a typical single-shot image of the spatial distribution of electrons, measured with an image plate (IP) stack, which was placed below the horizontal plane around the target rear in some shots. (b), a typical energy spectrum of the electrons. (c). a typical image of the THz spot on the target rear surface.

The energy spectra of electrons were detected using an electron spectrometer, consisting of a magnet with field strength of 0.1 T and an image plate (IP) detector. The spatial distribution of the electron bunch was recorded by a four-layered IP stack [24], which was wrapped with a 1mm-thick Fe filter to shield from the visible light, low-energy x-ray photons and ions. In order not to affect the electron spectrometer and Thompson ion spectrometer simultaneously, the IP stack was positioned 5 mm below the horizontal plane at a distance of 100 mm from the target. At the pump laser energy of ~60 J, the electron temperature was measured to be ~2 MeV, as shown in Fig. 1(b). A typical single-shot image of the spatial distribution of electrons measured on the second-layer IP is shown in Fig. 1(a). The electrons showed a double-peak angular distribution, primarily peaked along the rear target normal direction (0°), and with a sub-peak near the laser direction (−30°), implying that the electrons were accelerated via the combined actions of Brunel heating and $\mathbf{J} \times \mathbf{B}$.

By integrating the signals on IP and considering the finite spatial-energy spectral detection range of IP, the total electron bunch charge emitted from the target rear was estimated to be ~340 nC, corresponding to a peak current of ~0.2 MA.

Intense THz radiation was observed. At a laser energy of ~60 J, the THz energy in 0.12 sr at 75° was measured to be ~2.3 mJ (<20 THz). Spectral measurements, either with a set of low-pass filters or narrowband band-pass filters, showed that the THz radiation was low-frequency (<3 THz) dominated (Fig. 2a). The THz energy measurements at different directions show that the THz radiation is weaker near the rear target normal direction (Fig. 2b).

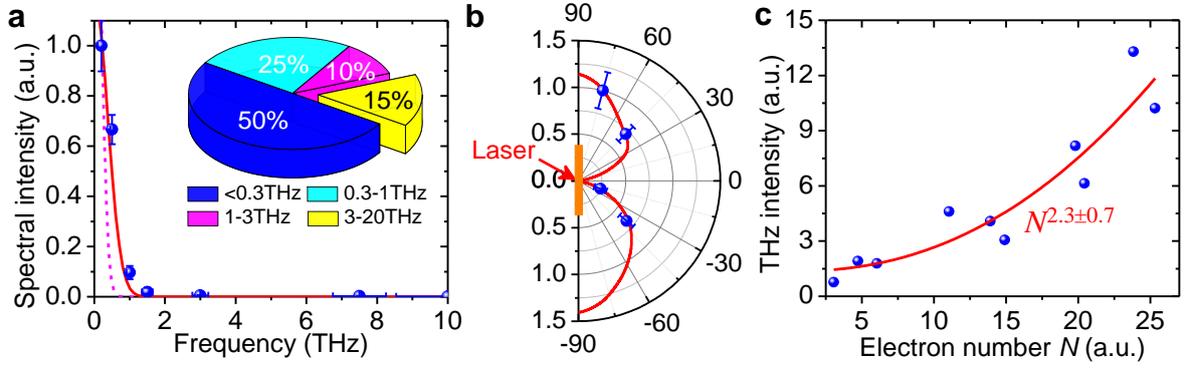

**Fig. 2.** Characterization of THz radiation. (a) THz spectra measured with band-pass filters (blue circles) and low-pass filters (inset). The solid red and dashed magenta curves show the theoretically estimated spectral range with and without taking into account the electron bunch divergence, respectively. (b) Measured (blue circles) and calculated (red curve) angular distributions of THz radiation. The red arrow and orange slab refer to the laser-target configuration. (c) Dependence of THz radiation on the measured electron number (blue circle) and the fit with a power-law function (red curve).

The coherent transition radiation (CTR) model [16,18] can explain the THz generation very well. For an electron bunch crossing a conductor-vacuum interface, the CTR energy $W_{CTR}$ emitted per unit angular frequency $d\omega$ and unit solid angle $d\Omega$ can be expressed as

$$\frac{d^2 W_{CTR}}{d\omega d\Omega} \propto N^2 \left( \left| \int d^3 \boldsymbol{p} g(\boldsymbol{p}) \xi_\parallel F(\omega) \cdot \boldsymbol{D} \right|^2 + \left| \int d^3 \boldsymbol{p} g(\boldsymbol{p}) \xi_\perp F(\omega) \cdot \boldsymbol{D} \right|^2 \right), \qquad (1)$$

where $N$ is the number of electrons, $g(\boldsymbol{p})$ is the normalized momentum distribution function, and $F(\omega)$ is the bunch form factor, defined as the square of the Fourier transform of the normalized electron bunch temporal profile. $\xi_\parallel$ and $\xi_\perp$ are the normalized amplitudes of electric fields in the radiation plane (defined by the target normal and the observation direction) and perpendicular to the radiation plane, respectively. *D* is the correction factor due to the finite target size (~3 mm in the experiment). Expressions for $\xi_\parallel, \xi_\perp$ and *D* can be found in Ref. [25].

To confirm that the observed THz radiation originates from the coherent emission of electrons, we measure the THz intensity as a function of electron number by varying the laser energy. A good agreement with the theoretical $N^2$ correlation is obtained, as shown in Fig. 2(c). The deviation from the perfect quadratic dependency may arise from the increase of electron temperature with the laser energy.

The bunch form factor, $F(\omega)$, serves as the major determinant of radiation spectra. For a collimated electron bunch with a Gaussian temporal distribution, $F(\omega)=\exp(-\omega^2\tau_0^2/2)$, where $\tau_0$ is the root-mean-square bunch duration. The duration of the laser-accelerated electron bunch is typically on the order of the laser pulse duration, which is ~1.5 ps in the experiment. Accordingly one can calculate the radiation spectra, as shown by the dashed magenta curve in Fig. 2(a). The experimentally measured spectrum shows a broader bandwidth than the theoretical one. This results from the rather large divergence angles of the electron bunch generated in the experiment. The finite bunch divergence will reduce the radiated energy, and broaden the radiation spectrum as well [26]. The relative spectral broadening can be estimated as $\delta\omega/\omega \sim \sqrt{\gamma_\perp} \sim 1$, where $\gamma_\perp \sim 1$ is the transverse electron kinetic energy normalized by electron static energy, for the measured electron bunch with a temperature of ~2 MeV and a divergence angle of ~28°. Taking into account the spectral broadening effect, the theoretical radiation spectrum is in good agreement with the measured, as shown by the red curve in Fig. 2(a).

By substituting the measured angular distribution and energy spectra of the electron bunch (inset a and b in Fig. 1) into Eq. 1, one can numerically calculate the angular distribution of CTR. The calculated angular distribution in the detection plane agrees well with the experimental measurements, as shown in Fig. 2(b).

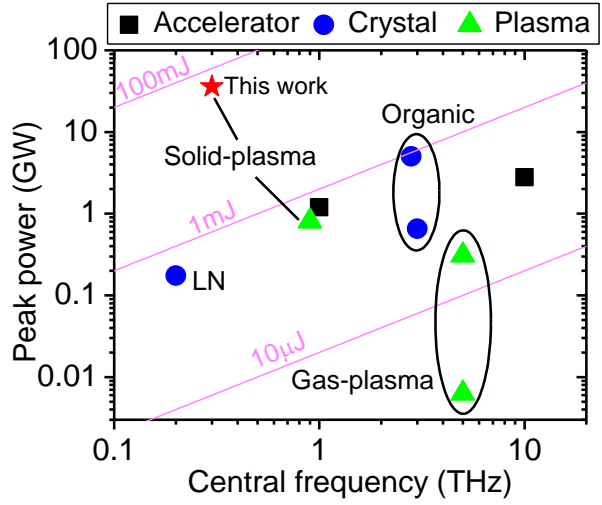

**Fig. 3.** Comparison of currently available high-peak-power THz sources. The data are referenced from previously reported typical results of THz sources based on conventional accelerators [10,11] (black squares), optical rectification from crystals (blue circles) like lithium niobate (LN) [12] and organic crystals [13,14], and gas [21,22] /solid-density plasmas [15] (green triangles). The red star represents the data presented in this paper. Magenta curves represent different energy ranges for half-cycle THz pulses.

The spatially integral of the calculated angular distribution indicates that the total energy emitted from the target rear surface is ~28 times the energy in 0.12 sr at 75°, which is measured to be 1.96 mJ within 3 THz in the experiment. Hence the total THz energy emitted from the target rear, at a pump laser energy of ~60 J, is determined to be ~55 mJ. This corresponds to a peak power of ~36 GW and a laser-THz energy conversion efficiency of ~0.1%. To our knowledge, this is the highest THz pulse energy and peak power reported so far (Fig. 3). Taking into account the THz source size of ~2 mm (Fig. 1c), the THz electric field, $E_{THz}$, at the target rear can reach ~3 GV/m. Although a higher THz peak field of 8.3 GV/m has been reported at multi-THz central frequencies and thus in a much smaller focal spot [14], a comparably low central frequency of ~0.3 THz here leads to a much higher ponderomotive potential $U_p$~100 keV, defined as the electron kinetic energy gained from the half-cycle THz pulse. The normalized vector potential, as a critical parameter to characterize the electromagnetic field strength, is evaluated as $a_0 = eE_{THz}/m_e c\omega_0$ ~0.9, where $\omega_0$ is the central frequency, $e$ and $m_e$ are the charge and mass of the electron, respectively, and $c$ is the speed of light. This almost approaches the realm of relativistic optics, which is not accessible on a half-cycle time scale previously.

In conclusion, we have experimentally demonstrated the generation of low-frequency (<3THz) THz pulses with a record high energy of over 50 mJ, exceeding other state-of-the-art THz sources by more than one order of magnitude. It is attributed mainly to the coherent transition radiation induced by the ultrahigh-current energetic electron bunch generated during relativistic ps laser-solid interactions. Such a THz source with ultra-high pulse energies reported here could enable the study of a fundamentally new physical domain of relativistic THz optics. Together with intrinsically synchronized energetic particles and photons generated concomitantly in laser-plasma interactions, new opportunities will be opened up in extremely nonlinear THz applications.

**Acknowledgements**

We would like to acknowledge the support and expertise of the staff of the Central Laser Facility. This work is supported by the National Basic Research Program of China (Grants No. 2013CBA01501), the National Nature Science Foundation of China (Grants No. 11520101003, and 11535001), the Strategic Priority Research Program of the Chinese Academy of Sciences (Grant No. XDB16010200 and XDB07030300), and the Newton UK grant. P.M. and D.N. acknowledge support from EPSRC (Grant numbers EP/R006202/1 and EP/K022415/1). G.L. acknowledges support from the National Postdoctoral Program for Innovative Talents (Grant No. BX201600106).